\documentclass[lettersize]{IEEEtran}
\usepackage{cite}
\usepackage{amsmath,amssymb,amsfonts, amsthm}
\usepackage{algorithmic}
\usepackage{graphicx}
\usepackage{textcomp}
\usepackage{xcolor}
\usepackage{booktabs}
\usepackage{tabularx}
\usepackage{algorithm}
\usepackage{algorithmic}
\usepackage{bbm}
\usepackage{bm}
\usepackage{bbold}
\usepackage{multirow}
\usepackage{float}
\usepackage{glossaries}
\usepackage{siunitx}
\usepackage{accents}
\newacronym{3gpp}{3GPP}{3rd Generation Partnership Project}    
\newacronym{6g}{6G}{sixth generation of mobile networks}
\newacronym{an}{AN}{anchor node}
\newacronym{awgn}{AWGN}{additive white Gaussian noise}
\newacronym{ber}{BER}{bit error rate}
\newacronym{bler}{BLER}{block error rate}
\newacronym{cdf}{CDF}{cumulative distribution function}
\newacronym{cml}{CML}{commercial microwave link}
\newacronym{iot}{IoT}{Internet of Things}
\newacronym{isac}{ISAC}{integrated sensing and communications}
\newacronym{ntn}{NTN}{Non-Terrestrial Network}
\newacronym{leo}{LEO}{Low Earth Orbit}
\newacronym{vleo}{VLEO}{Very Low Earth Orbit}
\newacronym{geo}{GEO}{Geosynchronous Earth Orbit}
\newacronym{isl}{ISL}{Inter-Satellite Link}
\newacronym{gsl}{GSL}{Ground-to-Satellite Link}
\newacronym{qos}{QoS}{Quality of Service}
\newacronym{ofdma}{OFDMA}{Orthogonal Frequency-Division Multiple Access}
\newacronym{b5g}{B5G}{5G and Beyond}
\newacronym{mimo}{MIMO}{Multiple-Input Multiple-Output}
\newacronym{mcs}{MCS}{modulation and coding scheme}
\newacronym{embb}{eMBB}{Enhanced Mobile Broadband}
\newacronym{ue}{UE}{User Equipment}
\newacronym{nr}{NR}{New Radio}
\newacronym{gap}{GAP}{Generalized Assignment Problem}
\newacronym{mgap}{MGAP}{Multi-Level Generalized Assignment Problem}
\newacronym{ml}{ML}{machine learning}
\newacronym{mac}{MAC}{medium access control}
\newacronym{csi}{CSI}{channel state information}    
\newacronym{ran}{RAN}{radio access network}
\newacronym{5g}{5G}{the 5th generation of mobile networks}
\newacronym{uav}{UAV}{unmanned aerial vehicle}
\newacronym{snr}{SNR}{signal-to-noise ratio}
\newacronym{ra}{RA}{resource allocation}
\newacronym{rssi}{RSSI}{received signal strength indicator}
\newacronym{rv}{RV}{random variable}
\newacronym{ul}{UL}{uplink}
\newacronym{dl}{DL}{downlink}
\newacronym{mle}{MLE}{maximum likelihood estimator}
\newacronym{crlb}{CRLB}{Cramér-Rao lower bound}
\newacronym{mse}{MSE}{mean-squared error}
\newacronym{nmse}{NMSE}{normalized mean-squared error}
\newacronym{ppp}{PPP}{Poisson point process}
\newacronym{milp}{MILP}{mixed-integer linear problem}
\newacronym{rtt}{RTT}{round-trip time}
\newacronym{noma}{NOMA}{non-orthogonal multiple access}
\newacronym{jmra}{JMRA}{joint matching and resource allocation}
\newacronym{dmrab}{DMRAB}{disjoint matching and resource allocation benchmark}
\newacronym{kpi}{KPI}{key performance indicator}
\newacronym{dtmc}{DTMC}{discrete-time Markov chain}
\newacronym{sca}{SCA}{successive convex approximation}
\newacronym{ngso}{NGSO}{non-geostationary orbit}
\newacronym{fr3}{FR3}{Frequency Range 3}
\newacronym{tn}{TN}{Terrestrial Network}
\usepackage[font=footnotesize]{caption}
\usepackage[font=scriptsize]{subcaption}
\usepackage{tikz}
\usepackage{pgfplots}
\newcommand{\set}{\mathcal}
\theoremstyle{definition}
\newtheorem{definition}{Definition}[]

\def\BibTeX{{\rm B\kern-.05em{\sc i\kern-.025em b}\kern-.08em
    T\kern-.1667em\lower.7ex\hbox{E}\kern-.125emX}}
\usepackage{balance}

\newcommand{\mc}[1]{\mathcal{#1}}   

\begin{document}
\title{ISAC-Powered Distributed Matching and \\  Resource Allocation in Multi-band NTN}
\author{Israel Leyva-Mayorga\IEEEauthorrefmark{1}, Shashi Raj Pandey\IEEEauthorrefmark{1}, Petar Popovski\IEEEauthorrefmark{1}, 
Fabio Saggese\IEEEauthorrefmark{2}, \\Beatriz Soret\IEEEauthorrefmark{3}\IEEEauthorrefmark{1}, and \v Cedomir Stefanovi\'c\IEEEauthorrefmark{1}\\
\IEEEauthorblockA{\IEEEauthorrefmark{1}Department of Electronic Systems, Aalborg University, Aalborg, Denmark (\{ilm, srp, petarp, bsa, cs\}@es.aau.dk)}\\
\IEEEauthorblockA{\IEEEauthorrefmark{2}Department of Information Engineering, University of Pisa, Italy (fabio.saggese@ing.unipi.it)}\\
\IEEEauthorblockA{\IEEEauthorrefmark{3}Telecommunications Research Institute, University of Malaga, Spain (bsa@uma.es)}
}

\maketitle
\begin{abstract}
Scalability is a major challenge in \gls{ngso} satellite networks due to the massive number of ground users sharing the limited sub-6 GHz spectrum.
Using K- and higher bands is a promising alternative to increase the accessible bandwidth, but these bands are subject to significant atmospheric attenuation, notably rainfall, which can lead to degraded performance and link outages. We present an \gls{isac}-powered framework for resilient and efficient operation of multi-band satellite networks. It is based on distributed mechanisms for atmospheric sensing, cell-to-satellite matching, and \gls{ra} in a 5G \gls{ntn} wide-area scenario with quasi-Earth fixed cells and a beam hopping mechanism. Results with a multi-layer multi-band constellation with satellites operating in the S- and K-bands demonstrate the benefits of our framework for \gls{isac}-powered multi-band systems, which achieves $73$\% higher throughput per user when compared to  single S- and K-band systems. 
\end{abstract}
\glsresetall

\section{Introduction}

Non-geostationary orbit (NGSO) satellite networks, particularly \gls{leo} constellations, have emerged as a critical technology for global connectivity, both as standalone systems and as complementary extensions to terrestrial 5G networks~\cite{leyva2020leo}. Nevertheless, scalability remains a fundamental challenge: the sheer number of potential ground users, coupled with severely limited bandwidth in conventional sub-6~GHz satellite bands, necessitates innovative approaches to resource management and optimization.

Operating satellite networks in millimeter-wave bands, such as the K-band and above, is a promising avenue to address bandwidth scarcity~\cite{SatBook}. The abundant spectrum  at these frequencies enables substantially greater data throughput capabilities compared to  lower satellite bands. However, this potential gain entails a critical limitation: signals in K-band frequencies and higher suffer severe attenuation due to atmospheric effects, particularly rainfall~\cite{itu}. The presence of water particles in the atmosphere causes pronounced signal degradation, and can lead to link outages, directly impacting link availability and reliability. This trade-off between increased capacity and atmospheric vulnerability necessitates mechanisms to sense and adapt to environmental conditions in real time.

The allocation of communication resources between satellite-cell pairs is fundamentally complex due to the dynamic nature of ground-to-satellite links and the diverse traffic demands~\cite{kisseleff2021RRM}. The problem becomes even more challenging in satellite networks with multiple layers (i.e., orbital shells and/or aerial and terrestrial nodes) and multiple frequency bands, as coordination between the nodes increases signaling overhead and  frequency allocation adds an extra dimension to the problem~\cite{Zhu25}. Thus, centralized optimization approaches face scalability challenges due to the exponential complexity of the underlying combinatorial problems, the large problem dimension, and the communication overhead required to exchange global state information between network elements~\cite{Leyva25}. Distributed solutions, on the other hand, may suffer from degraded performance. However, \gls{isac} offers a promising approach to achieve near-optimal performance without requiring network-wide coordination by providing real-time \gls{csi} to execute distributed \gls{ra} mechanisms.

This paper presents an \gls{isac}-powered framework to achieve a resilient and efficient operation of multi-band satellite networks  operating in 5G \gls{ntn} quasi-Earth fixed cell scenarios with beam hopping capabilities. Our approach enables the flexible coverage of ground cells by combining distributed mechanisms for atmospheric sensing and \gls{ra}, leveraging a preference-based matching algorithm informed by estimated channel conditions. 
We propose a deferred acceptance matching algorithm that relies on atmospheric sensing to construct preference lists for both cells and satellites, enabling a stable, low-complexity \gls{ra} solution that approaches the performance of centralized methods while maintaining the practicality of a distributed approach. 
The key contributions of this work are as follows.
\begin{enumerate}
    \item An introduction of a 5G \gls{ntn}-compliant frame structure based on \gls{ofdma}, specifying the resources for \gls{dl} communication, sensing, cell-to-satellite and resource allocation to support the proposed framework.

    \item An adaptation of the deferred acceptance algorithm to achieve a stable assignment of cells to serving satellites for 5G \gls{ntn} scenarios with beam hopping satellites. This algorithm operates in combination with a local \gls{ra} method and uses the \emph{quota} (i.e, the available resources at the satellites) to provide an upper bound for the computational complexity of \gls{ra} at the satellites.

    \item A thorough analysis of the sensing accuracy and communication performance in multi-layer and multi-band satellite constellations showing that, with our framework, constellations operating at the S- \emph{and} K-bands can achieve a $73$\% increase in per-user throughput when compared to an equivalent S-band only constellation.
\end{enumerate}

The rest of the paper is organized as follows. Sec.~\ref{sec:sys_model} presents the system model, Sec.~\ref{sec:methods} presents the frame structure, the matching and the \gls{ra} algorithms. Sec.~\ref{sec:results} presents the performance evaluation and Sec.~\ref{sec:conclusions} draws the conclusions.






\begin{figure}[t]
    \centering
    \subfloat[]{\includegraphics[width=0.88\columnwidth]{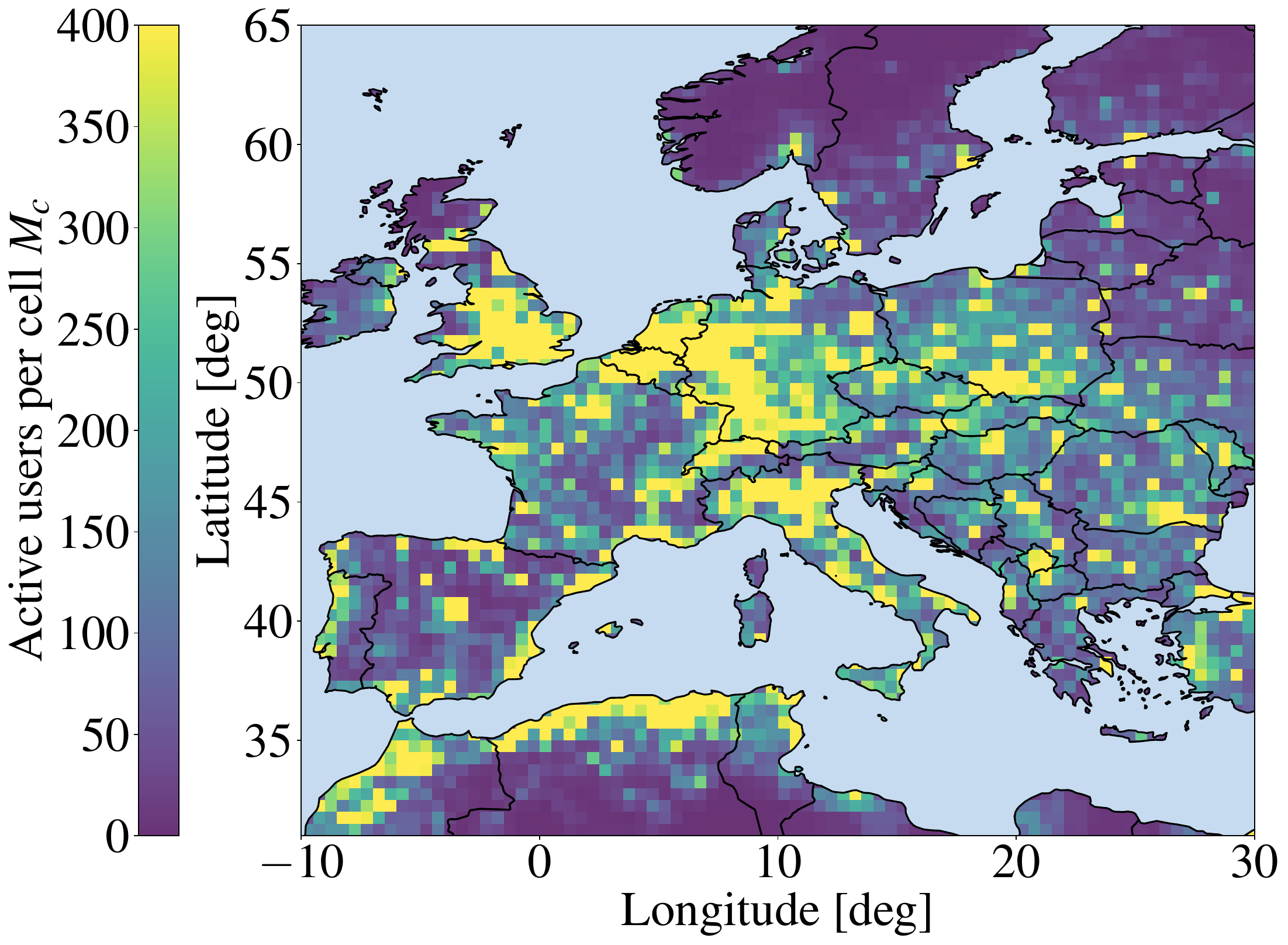}}\\
    \subfloat[]{\includegraphics[width=0.88\columnwidth]{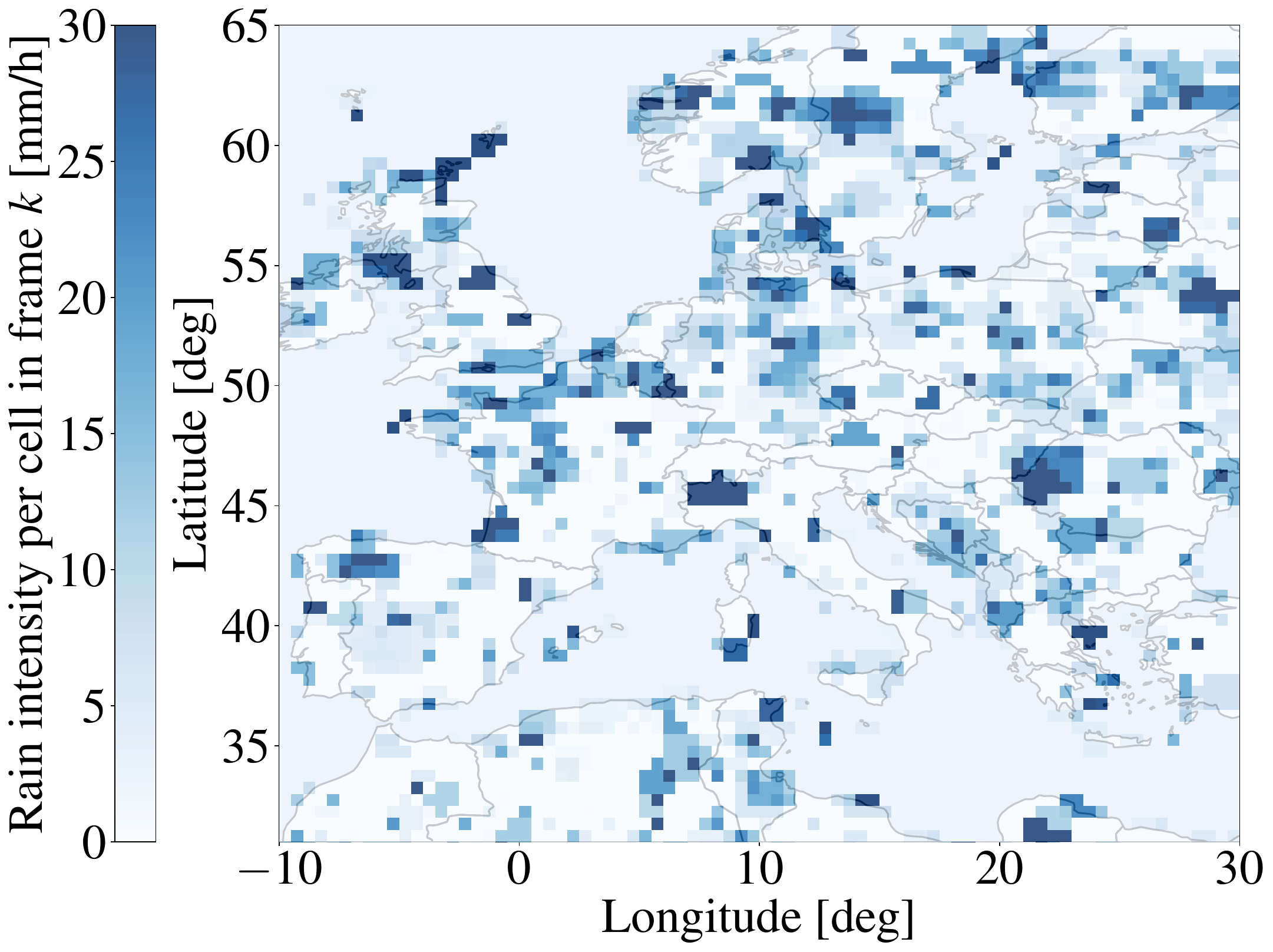}}
    \caption{Map with (a) active users and (b) rain intensity in the area of interest.}\vspace{-0.5em}
    \label{fig:users_map}
\end{figure}

\section{System model}
\label{sec:sys_model}
We consider a direct satellite-to-user \gls{dl} scenario, where a multi-layer satellite constellation with $S$ satellites serves $M$ active users in a pre-defined geographical region~\cite{TS38.300,Guidotti2018}. These users are organized into $C$ fixed and evenly-distributed geographical cells according to the quasi-Earth fixed cell scenario~\cite{3GPPTR38.821, TS38.300}, shown in Fig.~\ref{fig:users_map}. The set of cells is denoted as $\set{C}$ and the  number of active users in cell $c\in\mc{C}$ is $M_c$, which is a fraction $\alpha_c\in\left[0,1\right]$ of the cell population. 


The set of satellites forming the constellation $\mathcal{S}$ is  divided into disjoint subsets of satellites $\mathcal{S}_0, \mathcal{S}_1,\dotsc,\subseteq \set{S}$, which denote different orbital shells deployed at altitude $h_i$ and inclination angle $\delta_i$, each having its own characteristics. Each satellite $s \in \mathcal{S}_i$ has $\beta_s$ independent beams, operates at a given frequency band with the same carrier frequency $f_s$ and bandwidth $B_s$. All the satellites in the same orbital shell (i.e., $s\in\set{S}_i$) share the same $f_s=f_i$, $B_s=B_i$, $\beta_s=\beta_i$, and $\delta_s=\delta_i$. Finally, satellites operating at a K-band or higher carrier frequency can participate in atmospheric sensing, as their signals are attenuated by particles in the atmosphere. We denote this subset of \emph{sensing enabled satellites} as $\mc{S}_\mathrm{sens}\subseteq\mc{S}$.

We consider a discrete-time model for operation of the system and the evolution of the environment dynamics. Specifically, time is divided into \emph{system frames} (hereafter simply referred as frames) of duration $T$ and indexed by $k\in\{1,2,\dotsc\}$. At the beginning of each frame, the position of the satellites, the wireless channels, and the environmental conditions are updated, and they are assumed to remain static within the entire frame. The operation of the system is compliant with 5G \gls{nr} \gls{ntn} standards~\cite{3GPPTR38.821, 3GPPTS38.211}, where an integer number of \gls{ofdma} frames $N_T\in\mathbb{N}$, of duration $T_F$, are contained in each frame, so $T=N_TT_F$. Among these $N_T$ \gls{ofdma} frames, $N_C$ are reserved for \gls{dl} communication, $N_S$ are reserved for atmospheric sensing, and $N_\text{FB}$ are reserved for feedback after sensing, such that $N_T=N_C+N_S+N_\text{FB}$.


Let $d_{s,c}(k)$ be the cell-edge distance (i.e., the maximum distance between satellite $s$ and any point in cell $c$) at frame $k$ and $v_c$ is the speed of light. The free-space path loss between cell $c$ and a satellite $s\in\set{S}$ at the $k$-th frame is given by 
\begin{equation}
   \set{L}_{s,c}(k) = \left(4 \pi d_{s,c}(k) f_s\right)^2 v_c^{-2}. 
   \label{eq:loss}
\end{equation}

Next, let $P_s$ and $G_s$ respectively be the transmission power and the antenna gain for satellite $s$, and $G_\text{gnd}$ the antenna gain of the ground devices. Thus, the \gls{snr} from any user in cell $c$ to satellite $s$ at frame $k$ is at least
\begin{equation}
    \gamma_{s,c}(k)=\frac{P_s G_s G_\text{gnd}}{\set{L}_{s,c}(k) A_{s,c}(k)\,\varphi\,\sigma^2},
    \label{eq:snr}
\end{equation}
where $A_{s,c}(k)$ is the atmospheric attenuation for the link, $\varphi$ is the pointing loss, and $\sigma^2$ is the noise power at the receiver. While the methods and algorithms proposed in this paper are agnostic to the atmospheric attenuation model, which allows to consider diverse atmospheric phenomena~\cite{Zinevich2010, Jian2022compressive}, for the sake of simplicity, we assume the atmospheric attenuation is only generated by rain. Therefore, let $\varrho_c(k)$ be the rainfall intensity in mm/h for cell $c$ and $\tilde{d}_{s,c}(k)$ be the distance along the line-of sight path between the ground terminal in cell $c$ and satellite $s$ that traverses the rain layer~\cite{ITU-rainheight}. Then, the attenuation between satellite $s$ and cell $c$ during time slot $k$ is 
\begin{equation} \label{eq:atm-loss}
    10\log_{10}\left(A_{s,c}(k)\right) = a_s \, \left[\varrho_c\left(k\right)\right]^{b_s} \tilde{d}_{s,c}(k),
\end{equation}
where coefficients $a_s$ and $b_s$ depend on $f_s$ and the polarization of the signal transmitted by the satellite $s$~\cite{ITU-rain, Zhao2001}. 

Assuming that the time-frequency resources within each cell are distributed evenly among the $M_c$ active users, the instantaneous data rate from satellite $s$ to any user in cell $c$ is
\begin{equation}   
    \rho_{s,c}(k)  =
    \frac{B_s}{M_c}\log_2\left(1+\gamma_{s,c}(k)\right).
    \label{eq:achievable-rate}
\end{equation}

We consider the distribution of cells to satellites and of satellite beams to cells as a \emph{matching and \gls{ra} problem}.
The matching problem aims to define a subset of cells to be served by satellite $s$ during frame $k$. On the other hand, the \gls{ra} problem aims to define the beam hopping pattern for each frame $k$, given by the set $\set{X}(k)=\{x_{s,c}(k)\}_{s\in\set{S}, c\in\set{C}}$, where $x_{s,c}(k)\in\{0,\ldots, N_C\}$ denotes the number of \gls{ofdma} frames in which a beam from satellite $s$ provides \gls{dl} communication with the users in cell $c$ during frame $k$. Therefore, the average throughput of a user in cell $c$ served by satellite~$s$ during frame $k$ is
    $\overline{R}_{s,c}(k)=\rho_{s,c}(k)\,x_{s,c}(k)/N_T$.



\subsection{Atmospheric sensing}
\label{sec:atmo}
Atmospheric sensing is performed in a bi-static manner with satellites transmitting pilot signals of length $L_p$ symbols to \glspl{an} on ground. We assume that one \gls{an} is deployed per cell and that their positions are known. 
Furthermore, each sensing satellite $s\in\set{S}_\mathrm{sens}$ uses a different pilot of length $L_p\geq \left\lvert\set{S}_\mathrm{sens}\right\rvert$, so the pilots can be designed to be fully orthogonal (i.e., not interfering each other).

 Let $m_\ell$, $z_\ell$, and $y_{\ell,s,c}$ be the $\ell$-th transmitted symbol, the complex sampled and zero-mean \gls{awgn} of unit variance, and the received symbol at cell $c\in\set{C}_s$ from satellite $s\in\set{S}_\mathrm{sens}$ in frame $k$, respectively. Then, the $\ell$-th received pilot symbol for cell $c$ is
 \begin{IEEEeqnarray}{rCl}
     y_{\ell,s,c} &=& m_\ell\sqrt{\gamma_{s,c}(k)\sigma^2} + z_\ell\sqrt{\sigma^2}.
  \end{IEEEeqnarray}
 The unbiased \gls{mle} for the \gls{snr} in a complex-valued channel given a known pilot signal of length $L_p$ symbols without upsampling is used, is~\cite{Paluzzi2000}
\begin{equation} \label{eq:hatgamma}
    \hat{\gamma}_{s,c}(k)=\frac{\left(L_p-\frac{3}{2}\right)\left(\frac{1}{L_p}\sum_{i=1}^{L_p}\mathrm{Re}\left\{y_{i,s,c}^*\,m_i\right\}\right)^{\!\!2}}{\sum_{i=1}^{L_p}|y_{i,s,c}|^2\!-\!\frac{1}{L_p}\!\left(\sum_{i=1}^{L_p}\mathrm{Re}\left\{y_{i,s,c}^*\,m_i\right\}\!\right)^{\!\!2}}.
\end{equation}
Furthermore, the \gls{crlb} for the \gls{snr} estimator in an \gls{awgn} channel is given as~\cite{Paluzzi2000}
\begin{equation}
    \text{var}\left(\hat{\gamma}_{s,c}(k)\right)\geq \left(2\gamma_{s,c}(k) + \gamma^2_{s,c}(k)\right)/L_p.
    \label{eq:crb}
\end{equation}
Let $\gamma_{s,c}\left(k\mid \varrho_c=0\right)$ be the \gls{snr} of the link from satellite $s$ to cell $c$ calculated assuming no rain attenuation (i.e., a rain intensity of $0$ mm/h). The latter can be accurately calculated from the known satellite positions and used in combination with~\eqref{eq:crb} to formulate the following bias-corrected rain attenuation estimator:
\begin{equation} \label{eq:att-hat}
        \hat{A}_{s,c}(k) 
    =\frac{\gamma_{s,c}\left(k\mid \varrho_c=0\right)}{\hat{\gamma}_{s,c}(k)\left(1+1/L_p\right)+2/L_p}.
\end{equation}
Assuming the communication parameters of sensing satellites $s\in\set{S}_\text{sens}$ 
and $M_c$ are known, these are used to estimate the per-user rates for the links to the cells $c\in\set{C}_s(k)$ as
\begin{equation}   
    \hat{\rho}_{s,c}(k)  =
    \frac{B_s}{M_c}\log_2\left(1+\hat{\gamma}_{s,c}(k)\right).
    \label{eq:est_achievable-rate}
\end{equation}
Conversely, the \gls{snr} estimation and per-user rates for the satellites that do not participate in sensing $s\notin\set{S}_\text{sense}$ and their cells $c\in\set{C}_s(k)$ are calculated from the satellite positions assuming no rain attenuation as $\hat{\gamma}_{s,c}(k)=\gamma_{s,c}\left(k\mid \varrho_c=0\right)$. Finally, $\hat{\gamma}_{s,c}(k)=0$ for all $c\notin\set{C}_s(k)$ and $s\in\set{S}$.


\section{ISAC-powered framework}
\label{sec:methods}
We now present the framework for \gls{isac}-powered matching and \gls{ra}, which comprises the system \gls{mac} frame structure enabling the proposed solution, the distributed cell-to-satellite matching algorithm, and the formulation and solution of the local \gls{ra} problem.

\subsection{Frame structure}
\label{sec:frame}
\begin{figure}
    \centering
    \includegraphics[width=\columnwidth]{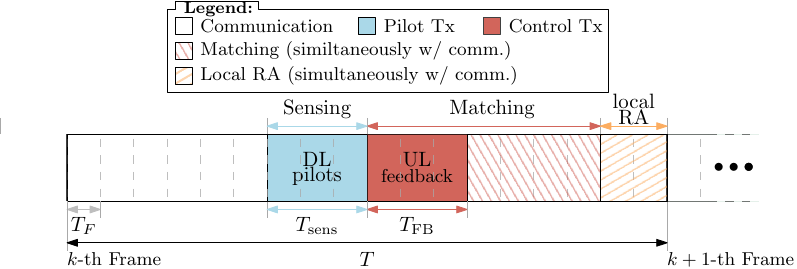}
    \caption{System frame designed for the proposed framework. A frame with $N_T$ \gls{ofdma} frames contains $N_C$ \gls{ofdma} frames for communications, $N_S$ for \gls{dl} sensing pilots and $N_\text{FB}$ for \gls{ul} priority lists transmission.}\vspace{-0.7em}
    \label{fig:frame}
\end{figure}

Fig.~\ref{fig:frame} shows the proposed \gls{ntn} frame-based structure designed to support sensing, satellite-to-cell matching, and \gls{ra}. The $N_T$ \gls{ofdma} frames in the overall system frame are divided into three groups: 1) $N_C$  communication frames, 2) $N_S$ sensing frames, and 3) $N_\text{FB}$ sensing feedback frames. Note that the \emph{matching} and \emph{local \gls{ra}} tasks do not require communication with the users and can be performed  simultaneously with \gls{dl} communication. 
The details on the sensing and sensing feedback are as follows. 

\subsubsection{Sensing}
\label{sec:sensing}
Sensing takes place by transmitting orthogonal pilots in the \gls{dl}, each pre-assigned to a single sensing enabled satellite $s\in\mc{S}_\mathrm{sens}$, such that the pilot length $L_p \ge |\mc{S}_\mathrm{sens}|$.

Considering the $s$-th satellite overall bandwidth of $B_s$~Hz, a symbol bandwidth of $\Delta^\text{sym}_s$~Hz and a duration of $T_s^\text{sym}$~s, the number of symbols that can be transmitted within a symbol duration is $N_s^\text{sym} = B_s/\Delta_s^\text{sym}$. Consequently, the minimum duration of a pilot of length $L_p$ symbols for satellite $s$ is
\begin{equation}
    T_s^\mathrm{pil}=\left\lceil\frac{L_p}{N_s^\text{sym}}\right\rceil=\left\lceil\frac{L_p\, \Delta_s^\text{sym}}{B_s}\right\rceil.
\end{equation}
Denoting by $\mc{C}_s(k)\subseteq\mc{C}$ the set of cells within the footprint of satellite $s$ at frame $k$, sending a single pilot to each cell in $\mc{C}_s(k)$ via beam hopping among the $\beta_s$ beams takes
\begin{equation}
    T^\text{sens}_s (k) = \frac{d^\text{max}_s}{v_c}+\left\lceil\frac{C_s(k)}{\beta_s}\right\rceil T_s^\mathrm{pil}\, T_s^\text{sym},
\end{equation}
where $ C_s(k) =|\mc{C}_s(k)|$ and $d^\mathrm{max}_s = \max_{c\in\mc{C}_s(k)} d_{s,c} (k)$ is the maximum satellite-edge distance in frame $k$.
With \gls{ofdma} frames of duration $T_F$, the number of frames that must be allocated to guarantee that the sensing can be completed is
\begin{IEEEeqnarray}{rCl}
    N_S 
    &=&
    \max_{s\in\mc{S}_\text{sens}}~\, \left\lceil T^\text{sens}_s(k)T_F^{-1} \right\rceil,
\end{IEEEeqnarray}
and the time for sensing in each frame is $T_\mathrm{sens} = N_S T_F$.

\subsubsection{Sensing feedback}
\label{sec:feedback}
Once the pilots have been delivered, the \glspl{an} estimates the \gls{snr} for each of the satellites within coverage, as described in Sec.~\ref{sec:atmo}, and generate the priority list, whose entries are tuples $(c,s,\hat{\rho}_{s,c}(k))$. The list is transmitted to the serving satellite, which acts as a \emph{broker} for the cells in the next execution of the matching algorithm, see Sec.~\ref{sec:matching}. To calculate the time needed to transmit the list, let $L_\text{tuple}$ be the length for the floating point representation of an entry $L_\text{list}$ be the number of entries in the  list. With a minimum rate for the \gls{ul} communication of $R_\text{FB}$~bps, the amount of \gls{ofdma} frames needed to transmit the priority list is
\begin{equation}
N_\text{FB}= \left\lceil\frac{L_\text{tuple}L_\text{list}} {R_\text{FB} } \, T_F^{-1}\right\rceil. 
\end{equation}
The time for sensing feedback in the frame is $T_\mathrm{FB} = N_\mathrm{FB} T_F$.


\subsection{Cell-to-satellite many-to-one matching}
\label{sec:matching}
Determining the set of cells to be served by the satellites at each frame $k$ is formulated as a bipartite matching game, which allows for capturing the preferences of both cells (i.e., players) and satellites (i.e., resources). We consider a many-to-one matching of cells to satellites, defined in the following.

\begin{definition} \label{def:m-o_matching}A many-to-one matching \mbox{$\mu(k) = \{  \mu_s(k), \mu_c(k)\}$} at frame $k$ is a mapping from the set $\mathcal{V}$ into the set of all subsets of  $\mathcal{V}$ such that the following conditions hold for each $c\in\mathcal{C}$ and $s\in\mathcal{S}\cup \{0\}$ with quota $q_s$ and $q_0=\infty$:
\begin{IEEEeqnarray}{C}
    \mu_s(k)\subset\mathcal{C} \text{ and } \mu_c(k)\subset\mathcal{S},\IEEEyessubnumber\label{matching_c1}\\
    |\mu_s(k)| \leq \min\left\{q_s, C_s(k)\right\},
    \IEEEyessubnumber\label{matching_c2}\\
    |\mu_c(k)| = 1, \IEEEyessubnumber\label{matching_c3}\\
    c\in\mu_s(k)\iff \mu_c(k)=\{s\}.\IEEEyessubnumber \label{matching_c4}
\end{IEEEeqnarray}
\end{definition}
The set $\{0\}$ is a virtual vertex set included to ensure stability, representing the lack of resources. The constraints~\eqref{matching_c2} and~\eqref{matching_c3} ensure that the matching does not exceed the available quota (i.e., resource capacity) at the satellites and that no two satellites are associated to the same cell, respectively.

The solution to this many-to-one matching is found through an adaptation of the deferred acceptance algorithm proposed by Galey and Shapley~\cite{gale1962college}, taking the form of a \emph{stable matching}, where neither the players nor the resources can change their choices to improve the solution. The solution is, thus, Pareto optimal~\cite{gale1962college, Gu15}, in scenarios with no interference. The matching game is played iteratively as follows:
\begin{enumerate}
    \item Each player and resource generates a preference list based on an individual metric.
    \item The players propose to be matched to its preferred resource and deletes it from the preference list.
    \item Each resource collects the proposals from the players and chooses a subset of them based on its own preference and \emph{quota}. The rest of the players are rejected.
    \item The algorithm terminates if all players are matched or all the preference lists of the remaining players are empty. Otherwise, steps 2, 3, and 4 are performed again.
\end{enumerate}

We use the estimation of the \gls{snr} $\hat{\gamma}_{s,c}(k)$ to characterize the preference relation $\succeq_c$ of the cells over the set of satellites and the estimation of the achievable instantaneous rates $\hat{\rho}_{s,c}(k)$ to characterize the preference relation $\succeq_s$ of the satellites. Thus, we define the preference lists for cells and satellites as follows.


\begin{definition}\label{def:pref-list-cell}[Preference list for user cells]
A preference list for vertex $c\in\set{C}$ at frame $k$ is the ordered set of vertices $\mathcal{W}_{c,k}=(w_1,w_2,\dotsc)\subseteq\set{S}$ such that $\hat{\rho}_{w_j,c}(k)\succeq_c \hat{\rho}_{w_{j+1},c}(k)$.
\end{definition}

\begin{definition}\label{def:pref-list-sats}[Preference list for satellites]
A preference list for vertex $s\in\set{S}$ at frame $k$ is the ordered set of vertices $\mathcal{W}_{s,k}=(w_1,w_2,\dotsc)\subseteq\set{C}$ such that $\hat{\rho}_{s,w_j}(k)\succeq_s \hat{\rho}_{s,w_{j+1}}(k)$.
\end{definition}



In the original deferred acceptance algorithm~\cite{gale1962college}, a stable matching is achieved with the players sending requests directly to the resources according to their priority lists. However, such implementation would be greatly inefficient in our satellite scenario due to the use \emph{beam hopping} mechanisms to illuminate the cells $\set{C}$ with directive beams. Therefore, our implementation proposes using the serving satellite $s=\mu_c(k)$ at frame $k$ as a \emph{broker} for all cells $c\in\mu_s(k)$, which receives $\set{W}_{c,k}$ for all $c\in\mu_s(k)$ and relays them to the corresponding satellites until a stable matching is achieved for frame $k+1$. Therefore, the broker satellite in frame $k$ is always defined by the matching in the previous frame, i.e., $\mu_c(k-1)$.

Below, we describe the proposed \gls{isac}-powered version of the deferred acceptance algorithm for cell-to-satellite matching, which integrates sensing to obtain the \gls{csi} from the satellite-cell links at each time slot $k$. 
\begin{enumerate}
    \item Each satellite $s\in\set{S}_\text{sens}$ generates a beam hopping schedule for sensing, which includes all the cells within its individual footprint $\set{C}_s(k)$. The schedules are transmitted to the cells $c\in\mu_s(k)$ by their broker satellite $s=\mu_c(k)$.
    \item The satellites $s\in\set{S}_\text{sens}$ illuminate the $c\in\set{C}_s(k)$ according to the schedule to transmit a pilot of length $L_p$ to each \gls{an}, as described in Sec.~\ref{sec:sensing}. 
    \item The \glspl{an} for all $ c\in\set{C}_s(k)$ and $s\in\set{S}_\text{sens}$ receive the pilots and estimate the \gls{snr} $\hat{\gamma}_{s,c}(k)$ through~\eqref{eq:hatgamma}.
    \item Each cell $c\in\set{C}$ generates and transmits its preference list $\set{W}_{c,k}$  to its broker satellite $\mu_c(k)$, including the tuples $(c,w_j,\hat{\rho}_{w_j,c}(k))$ $\forall w_j\in\set{W}_{c,k}$ as described in Sec.~\ref{sec:feedback}. 
    \item After collecting the preferences, each broker satellite $s$ repeats the following process until $\set{W}_{c,k}=\emptyset$ or $\left\lvert\mu_{w_j}(k+1)\right\rvert=q_s$ for all $w_j\in\set{W}_{c,k}$ and $c\in\mu_s(k)$.
    \begin{itemize}
    \item The broker satellite $s=\mu_c(k)$ sends a \emph{connection request message} to each the satellite at top of the preference lists $\set{W}_{c,k}$ of all $c\in\mu_s(k)$, containing the tuple $(c,w_1,\hat{\rho}_{w_1,c}(k))$. 
    \item Each satellite $s\in\set{S}$ collects the connection requests from the cells into the set $\mu_s^\text{req}$ and adds a subset of the requesting cells $\mu_s^\text{acc}\subseteq\mu_s^\text{req}$, selected from the top of its preference list $\set{W}_{s,k}$, to its matching $\mu_s(k+1) = \mu_s(k+1) \cup\mu_s^\text{acc}$, such that $\left\lvert\mu_s(k+1)\right\rvert\leq q_s$.  The cells $\mu_s^\text{req}\setminus \mu_s^\text{acc}$ are rejected.
    
    \item  The rejected and accepted cells are communicated to the broker satellites, which update the matching  $\mu_c(k+1)=\{s\}$ for all $c,s$ such that $c\in\mu_s(k+1)$. 
    
    \item The broker satellites remove the preference lists for all $c\in\mu_s(k)$ in $\mu_s^\text{acc}$; for all the rejected cells, top preferences $w_1$ are updated with the second element in the lists and the process repeats from step 5.
\end{itemize}
\end{enumerate}
Once the matching algorithm terminates, the matching $\mu(k+1)$ is communicated to the corresponding cells and satellites, so each $s\in\set{S}$ perform a local \gls{ra} to allocate its $\beta_s$ beams among the cells $\mu_s(k+1)$ in the next frame, as described below.
\vspace{-1em}



\subsection{Local resource allocation (RA)}
In the proposed distributed implementation, each satellite $s\in\mc{S}$ locally perform the allocation of its resources, $\beta_s$ and $N_C$, for the cells selected by the matching algorithm described above, i.e.,  $\forall c\in\mu_s(k+1)$.
To attain a proportional fair allocation of resources, we formulate the local resource allocation problem for satellite $s$ as follows.
\begin{IEEEeqnarray}{CCll} \label{eq:dist_opt_problem}
\mathcal{P}_s\!: \max_{\set{X}(k+1)}&~&\IEEEeqnarraymulticol{2}{l}{\hspace{-0.5em} \sum_{c\in\mu_s(k+1)}\hspace{-1em} M_c\log\!\left(1+\frac{\hat{\rho}_{s,c}(k)x_{s,c}(k+1)}{N_T}\right)\!\!,}\IEEEyesnumber*\IEEEeqnarraynumspace\label{eq:opt_problem_dist_objective}\\
\text{s.t.}&&\hspace{-6mm}
x_{s,c}(k+1)\in\{0,1,\dotsc, N_C\}, \,&\forall c\in\mu_s(k+1)\IEEEyessubnumber*\IEEEeqnarraynumspace\\[0.5em]
&&\hspace{-6mm}\displaystyle \sum_{c \in \mu_s(k)}x_{s,c}(k+1)\leq N_C \beta_s,\IEEEyessubnumber*
\end{IEEEeqnarray}
We solve $\mathcal{P}_s$ by performing a continuous relaxation of the integer variable $x_{s,c}(k+1)$, which transforms $\mathcal{P}_s$ into a convex problem, which can be efficiently solved using widely-available interior point methods. Once the continuous solution is obtained, it is discretized and projected to map it back into the original feasible set of $\mathcal{P}_s$.
The complexity of solving  $\mathcal{P}_s$ through this approach is determined by the complexity of the interior point method. The optimization is performed over the cells included in the matching $\mu_s(k+1)$ and $\left\lvert\mu_s(k)\right\rvert\leq \min\left\{q_s, C_s(k)\right\}$, $\forall k$. Thus, in the typical regime where a satellite $s$' footprint covers more cells than their quota, $C_s(k)>q_s$, the complexity of solving the \gls{ra} is $\set{O}\left(q_s^3\right)$. 

\subsection{Benchmarks}
We compare the performance of the proposed distributed framework to a centralized joint matching and resource allocation benchmark (CB) solution~\cite{Leyva25}:\vspace{-0.5em}
\begin{IEEEeqnarray}{CCll} 
\label{eq:opt_problem}
\mathcal{P}_\text{CB}\!:\hspace{-0.4em} \max_{\set{X}(k+1)}&~&\IEEEeqnarraymulticol{2}{l}{\hspace{-0.3em}\sum_{c\in\mc{C}} M_c\log\!\left(\!1+\hspace{-1.2em}\sum_{s\in\mu_c(k+1)}\hspace{-1em}\!\frac{\hat{\rho}_{s,c}(k)x_{s,c}(k+1)}{N_T}\!\right)\!\!,}\IEEEyesnumber*\IEEEeqnarraynumspace\label{eq:opt_problem_objective}\\[0.2em]
\text{s.t.}&&
x_{s,c}(k+1)\in\{0,1,\dotsc, N_C\}, ~&\forall s,c\IEEEyessubnumber*\\[0.5em]
&&\displaystyle \sum_{c \in \mathcal{C}}x_{s,c}(k+1)\leq N_C \beta_s, &  \forall s \in \mathcal{S},\IEEEyessubnumber*\\
&&\hspace{-1em}\displaystyle\sum_{s \in \mu_c(k+1)} 
\mathbb{1}\left( x_{s,c}\left(k+1\right) >0 \right) \le 1, & \forall c \in 
\mathcal{C},~\IEEEyessubnumber*\IEEEeqnarraynumspace\label{eq:opt_problem_c3}
\end{IEEEeqnarray}
where $\mathbb{1}(\cdot)$ is the indicator function. Constraint~\eqref{eq:opt_problem_c3} enforces the many-to-one matching, where each cell can be served by one satellite only, such that $\left\lvert\mu_c(k+1)\right\rvert\leq1$. Problem $\set{P}_\text{CB}$ is solved using the method of multipliers with $n_\text{iter}$ iterations, which has a complexity $\set{O}\left(S(k)^3C^3n_\text{iter}\right)$.

In addition, we compare the performance of our framework to a lower bound in performance with no sensing satellites (i.e., $\set{S}_\text{sense}=\emptyset$) and to an upper bound in performance that assumes perfect \gls{csi} with no sensing overhead. 

\section{Results}
\label{sec:results}
We now turn to the performance evaluation of a multi-layer and multi-band satellite constellation with the proposed \gls{isac}-powered matching and distributed \gls{ra} framework. The default parameter settings for the performed simulations are listed in Table~\ref{tab:sim_params}, which consider S-band satellites in \gls{leo} and K-band satellites in \gls{vleo}, and a discrete-time Markov model for clustered rain~\cite{Leyva25}. The area of interest on the surface of the Earth, illustrated in Fig.~\ref{fig:users_map}, is bounded by the south west (SW) coordinate $(-10^\circ, 30^\circ)$  and north east (NE) coordinate $(30^\circ, 65^\circ)$, containing $3960$ cells and an average of $98.4$ satellites per frame. The maximum number of cells within the coverage of individual \gls{leo} and \gls{vleo} satellites are $1212$ and $194$, respectively. Further, we assume that $R_\text{FB}$ is set to complete the \gls{ul} feedback in $N_\text{FB}=1$ \gls{ofdma} frame.
\begin{table}[t]
\renewcommand{\arraystretch}{1.1}
\centering
\caption{Simulation parameters}
\begin{tabular}{@{}lccc@{}}
    \toprule \textbf{Parameter} & \textbf{Symbol} & \multicolumn{2}{c}{\textbf{Value}} \\\midrule 
   \textbf{Orbital shells} &$i$& LEO &VLEO\\\midrule
    Center frequency [GHz] & $f_i$ & $2$ & $20$\\
    Bandwidth [MHz]  & $B_i$ & $30$ & $400$\\
     Satellite antenna gain [dBi] & $G_{s,c}$ & $30$ & $38.5$\\
    Total number of satellites & $S_i$ & $720$ & $1584$\\
     Number of orbital planes & $O_i$  & $36$& $72$\\
     Altitude of deployment [km] & $h_i$  & $570$& $200$\\
     Inclination [deg] & $\delta_i$  & $70$& $53$\\
     Transmission power [W] &$P_i$ & $75$ & $75$\\ 
     Pointing loss [dB]  & $\varphi_\text{dB}$ & $0.3$ & $0.3$\\
     Minimum elevation angle [deg] & $\eta_\text{min}$ & $25$ & $25$\\ 
     Number of beams per satellite & $\beta_i$ & $19$ & $19$\\
     Sensing symbol duration [$\mu$s]& $T_i^\text{sym}$ & N/A & $71.35$\\ 
     Sensing symbol bandwidth [kHz] & $\Delta_i^\text{sym}$ & N/A & $15$\\ 
    \midrule
    \textbf{Ground segment}\\
    \midrule
    Number of cells in the area & $C$ & \multicolumn{2}{c}{$3960$}\\
    Antenna gain [dBi] & $G_\text{gnd}$& \multicolumn{2}{c}{$0$}\\
    Noise spectral density [dBm/Hz] & $N_{0,\text{dB}}$ & \multicolumn{2}{c}{$-176.31$}\\
    Population per cell $c$ & $M^\text{max}_c$ & \multicolumn{2}{c}{From~\cite{CIESIN}}\\
    Ratio of active users & $\alpha_c$ & \multicolumn{2}{c}{$0.001$}\\
\midrule 
\textbf{Rain parameters}\\\midrule
     Storm intensity (PPP) [storms/km$^2$] & $\lambda_\text{rain}$ &\multicolumn{2}{c}{$8.4\times10^{-4}$}\\
     Rain height [km] & $h_r$ & \multicolumn{2}{c}{$4$}\\
     Mean rain intensity & $\overline{\varrho}$ & \multicolumn{2}{c}{$8.77$}\\
     Mean rain cell diameter [km] & $d_\text{rain}$ & \multicolumn{2}{c}{$50$}\\
     Mean duration of a rain episode [h] & $\varepsilon$ & \multicolumn{2}{c}{$1.886$}\\
     Mean period between rain episodes [h] & $\tau$ & \multicolumn{2}{c}{$5.376$}\\
     \midrule
\textbf{Frame structure}\\
\midrule
    Frame duration [s] & $T_F$ &\multicolumn{2}{c}{$10$}\\
     \gls{ofdma} frame duration [ms] & $T$ & \multicolumn{2}{c}{$10$}\\
     Pilot length [symbols] & $L_p$ & \multicolumn{2}{c}{$256$}\\ 
     
     \bottomrule
\end{tabular}
\label{tab:sim_params}
\vspace{-10pt}
\end{table}

To promote a fair matching in line with $\set{P}_s$ and $\set{P}_\text{CB}$, we define the cell and satellite preferences, respectively, as\vspace{-0.2em}
\begin{equation}
    \hat{\rho}_{w_j,c}(k)\succeq_c \hat{\rho}_{w_{j+1},c}(k)\iff\hat{\rho}_{w_j,c}(k)\geq\hat{\rho}_{w_{j+1},c}(k).
    \end{equation}
\begin{equation}
    \hat{\rho}_{s,w_j}(k)\succeq_s \hat{\rho}_{s,w_{j+1}}(k)\iff\hat{\rho}_{s,w_j}(k)\leq\hat{\rho}_{s,w_{j+1}}(k).
\end{equation}
Hence, using Definitions~\ref{def:pref-list-cell} and~\ref{def:pref-list-sats}, the utility of the cells and satellites are calculated from the true achievable rates as $u_c(\mu) =  \rho_{s,c}(k), \text{ s.t., } s\in \mu_c(k)$  and
$u_s(\mu) = \sum_{c\in\mu_s(k)} 1/\rho_{s,c}(k)$.

\begin{figure}
    \centering
    \includegraphics{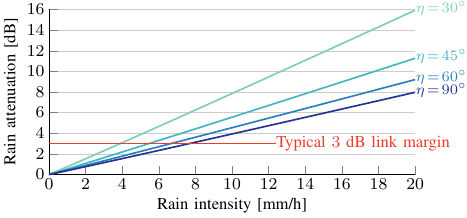}
    \caption{Attenuation due to rain for K-band satellites as a function of the rain intensity in mm/h for diverse elevation angles $\eta$.}
    \vspace{-1em}
    \label{fig:attenuation}
\end{figure}

As a starting point, Fig.~\ref{fig:attenuation} presents the rain attenuation for the signals of the \gls{vleo} satellites for different rain intensities and satellite elevation angles $\eta$. Evidently, typical link margins to compensate for the uncertainty of atmospheric attenuation are insufficient for rain intensities above $6$~mm/h. Furthermore, the rain attenuation increases significantly as $\eta$ decreases.  

\begin{figure}
    \centering
\includegraphics{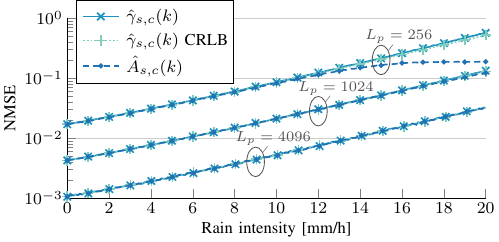}
    \caption{Normalized Mean Squared Error (NMSE) for the estimation of the SNR $\hat{\gamma}_{s,c}(k)$, its CRLB, and the rain attenuation $\hat{A}_{s,c}(k)$ for $\eta = 30$~deg and $L_p\in\{256, 1024, 4096\}$ as a function of the rain intensity in mm/h.}\vspace{-1.2em}
    \label{fig:nmse}
\end{figure}

Next, we present the \gls{nmse} for the \gls{snr} estimator $\hat{\gamma}_{s,c}(k)$, its \gls{crlb}, and the rain attenuation estimator $\hat{A}_{s,c}(k)$  as a function of the rain intensity $\varrho_c(k)$ and the pilot length $L_p\in\{256, 1024, 4096\}$ for a satellite elevation angle $\eta=30$~deg in Fig.~\ref{fig:nmse}. The values $L_p\in\{256, 1024\}$  allow to assign individual orthogonal pilots to all the satellites in the area of interest, while $L_p=4096$ allows to assign individual orthogonal pilots to all the satellites in the constellation. Clearly, the error for these estimators decreases as $L_p$ increases, and the error increases as the rain intensity increases due the increase in attenuation. Notably, $\hat{\gamma}_{s,c}(k)$ is extremely close to the \gls{crlb}, which indicates near-optimal estimation performance. Also, the estimator $\hat{A}_{s,c}(k)$ achieves a lower \gls{nmse} than $\hat{\gamma}_{s,c}(k)$ for all the illustrated rain intensities, achieving \glspl{nmse} below $10^{-2}$ with $L_p=1024$ and $L_p=4096$ for rain intensities below $6$ and $14$ mm/h, respectively.  

\begin{figure}
    \centering
    
    \includegraphics{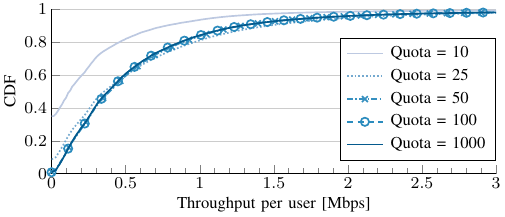}
    \caption{Per-user throughput \gls{cdf} for $L_p=256$ symbols and different quotas.}\vspace{-1em}
    \label{fig:throughput_quota}
\end{figure}

After validating the accuracy of the estimators, Fig.~\ref{fig:throughput_quota} presents the impact of diverse quota values at the satellites $q_s$ to the \gls{cdf} of the per-user throughput for the default pilot length $L_p=256$. Evidently, setting $q_s\geq50$ leads to similar per-user throughput and also to similar average cell and satellite utilities. In addition, setting $q_s\geq100$ leads to $100\%$ of the users being served by a satellite (i.e., $\left\lvert\mu_c(k)\right\rvert=1,\, \forall c\in\set{C}$) and to $99.1$\% of the maximum cell and satellite average utilities obtained with $q_s=1000$. Besides, further increasing $q_s$ has only a minimal impact on average per-user throughput, as a $0.2$\% decrease was observed with $q_s=1000$ when compared to $q_s=100$. Nevertheless, setting a high $q_s$ has a negative impact on the computational complexity of the \gls{ra}, as satellites were matched to up to $188$ cells for $q_s=1000$. On the other hand, using $q_s=25$ leads to a slightly higher per-user throughput than with $q_s\geq50$ for quantiles $\geq0.6$, but also leads to a $47$\% decrease in the average cell utility and to $12.4\%$ of users not having a serving satellite. The latter is a critical failure, since the users without a  broker satellite cannot perform the matching process for the next time slot. Further decreasing $q_s$ to $10$ increases the fraction of users without serving satellite to $35\%$ and decreases the cell and satellite average utilities by $86$\% and $34$\%, respectively. To serve all users while bounding the computational complexity to solve $\set{P}_s$, we set $q_s=100$ for the remainder of the paper.

\begin{figure}
    \centering
    \includegraphics{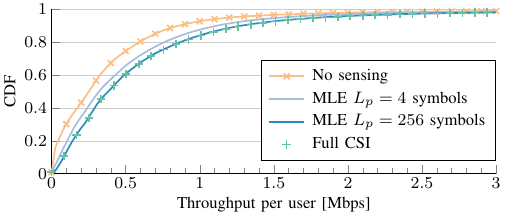}
    \caption{Per-user throughput for different $L_p\in\{4,256\}$, along with benchmarks with no sensing and full \gls{csi}.}\vspace{-0.5em}
    \label{fig:throughput_L_p}
\end{figure}

Next, Fig.~\ref{fig:throughput_L_p} illustrates the impact of the pilot length $L_p$ on communication performance, including the benchmarks with no sensing and with full \gls{csi}.
Setting $L_p=256$ leads to $99.4$\% of the average per-user throughput with full \gls{csi}, 
which practically reaches the upper bound with a moderate pilot length, and increases the average per user throughput by $49.5$\% when compared to no sensing. Besides, using extremely long sensing pilots  can decrease communication performance. For example, setting $L_p=2^{22}$ requires $N_S=11$ frames for sensing and, consequently, reduces the average per-user throughput by $1.3$\%.  
Conversely, setting $L_p=4$ only leads to $84.6$\% of the average per-user throughput  achieved by the full \gls{csi} benchmark. Therefore, we set $L_p = 256$ in the following.	

\begin{figure}
    \centering
    \includegraphics{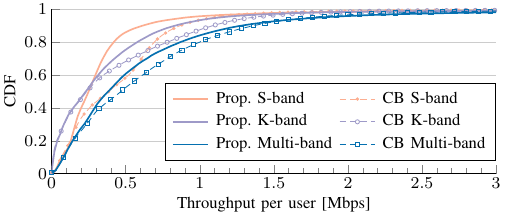}
    \caption{Per-user throughput CDF for different band allocations with the proposed framework (Prop.) and a centralized based solution (CB) benchmark.}\vspace{-1em}
\label{fig:throughput_constellations}
\end{figure}

Once adequate parameter settings for $q_s$ and $L_p$ have been identified, we compare the performance of the considered multi-band constellation and the proposed \gls{isac}-powered matching and distributed \gls{ra} framework (Prop.) to 1) single-band constellations and 2) the centralized benchmark (CB) from our previous work~\cite{Leyva25} in Fig.~\ref{fig:throughput_constellations}. Clearly, the multi-band constellation achieves the best performance, as it is able to exploit both the resilience of the S-band satellites to the environmental conditions and the increased bandwidth of K-band satellites. The use of S-band constellation only leads to lower per-user rates, as a consequence of the limited bandwidth. Further, the proposed method achieves similar per-user rates as the CB at the lowest (i.e., $\leq0.4$) and highest (i.e., $\geq 0.95$) quantiles. Hence, it is only in the mid-quantiles $[0.4,0.95]$ that the CB outperforms the proposed method. The proposed method with the multi-band constellation outperforms both S-band and K-band  constellations with the CB  in all per-user rate quantiles.

\begin{figure}[t]
    \centering
    \includegraphics{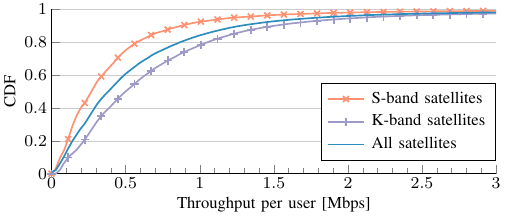}
    \caption{Per-user throughput CDF for the S-band \gls{leo} and K-band \gls{vleo} satellites in the multi-band constellation with the proposed framework.}\vspace{-1em}
\label{fig:throughput_per_band}
\end{figure}

Finally, we present the per-user throughput \gls{cdf} for the different serving S-band \gls{leo} and K-band \gls{vleo} satellites, as well as for all the serving satellites in Fig.~\ref{fig:throughput_per_band}. These results show that both types of satellites achieve comparable per-user throughput, which indicates that the configuration of the constellation is adequately balanced. 

\section{Conclusion}
\label{sec:conclusions}
We presented a framework for \gls{isac}-powered distributed matching and resource allocation, which includes the frame structure and the mechanisms for sensing, cell-to-satellite matching and resource allocation in multi-band satellite constellations. In a 5G-\gls{ntn} compliant scenario, a satellite constellation implementing our framework with satellites in S- and K-bands achieved $73$\% higher throughput per user when compared to an S-band constellation with similar deployment characteristics. Furthermore, we observed that $256$ pilot symbols for sensing lead to a $33$\% increase in per-user throughput when compared with no sensing and to $>99$\% of the upper bound in performance, achieved with perfect \gls{csi} and no sensing overhead. Future work includes extending the proposed sensing framework to fast fading channels.

\section{Acknowledgement}
This work was supported, in part, by the Velux Foundation, Denmark, through the Villum Investigator Grant WATER, nr. 37793 and by Danmarks Frie Forskningsfond (DFF) Project “3D-Twin” under Grant 10.46540/4264-00153B. The work of F. Saggese was supported by the Horizon Europe MSCA Postdoctoral Fellowships with Grant~101204088. The work of B. Soret was supported by the Spanish Ministry of Science and Innovation under Grant PID2022-136269OB-I00.
\vspace{-0.5em}
\bibliographystyle{IEEEtran}
\bibliography{bib.bib}
\end{document}